\documentstyle[12pt]{article}
     \begin{document}
      \vskip 2. true cm
      \begin{center}
      {\large \bf On the spatial density matrix for the centre
      of mass of a one dimensional perfect gas} \\

      \vskip 1.4 true cm
        B. Carazza \\
       \vskip .1 true cm
\par 
       {\it Dipartimento di Fisica dell' Universit\`a, 
       viale delle Scienze, I43100 Parma, Italy}\\
        {\it INFN Sezione di Cagliari, Italy}\\  
      \end{center}
\par \vskip 2mm
\begin{center} {\bf  Abstract} \end{center}
 \vskip 1mm
\begin{quote}
  We examine the reduced density matrix of the centre of mass
  on position basis considering a one-dimensional system of $N$ 
  non-interacting distinguishable particles in a infinitely deep square
  potential well. We find a class of pure states of the system for
  which the off-diagonal elements of the matrix above go to zero
  as $N$ increases. This property holds too for the state vectors
  which are factorized in the single particle wave functions.
  In this last case, if the average energy of each particle 
  is less than a common bound, the diagonal elements
  are distributed according to the normal law with a mean square deviation
  which becomes smaller and smaller as $N$ increases towards infinity.
  Therefore when the state vectors are of the type considered
  we cannot experience spatial superpositions
  of the  centre of mass and we may conclude that
  position is a preferred basis for the collective variable.
\end{quote}
\vskip 1.5 true cm
Key words: Macroscopic states, spatial macrosuperpositions,
 reduced density matrix, preferred basis, Central Limit Theorem.
 \newpage
      \noindent{\bf 1. INTRODUCTION}
      \vskip .6 true cm
 \par
      The old problem of how to derive the classical behaviour
      of macroscopic bodies from quantum principles now seems
      much closer to solution thanks to spontaneous localisation
      theories of the GRW type\cite{ghi} or by reference to
      decoherence which occurs for a system interacting with
      an environment\cite{omn}.
 \par
      In the case of a macroscopic body, decoherence is believed
      to suppress the off-diagonal elements of the spatial 
      reduced matrix of the collective variables, such as the
      centre of mass, irrespective of initial conditions. 
      This is equivalent to saying that we cannot
      experience spatial macrosuperpositions of the centre of
      mass and this explains why we do not encounter states
      of this kind when looking at everyday objects.
      Moreover, as long as we limit ourselves to considering spatial
      distributions of the collective variable and the observable
      quantities linked to this, we can interpret the reduced 
      density matrix after dechoerence in terms of classical
      probabilities.
  \par   
      Clearly if pure states exist for a closed macroscopic
      system, corresponding to a reduced density matrix for
      the centre of mass that is diagonal on the position
      basis, they will give the same conclusions for the
      description of the collective behaviour. 
      This has led us to identify possible pure states of 
      a system of many particles in an external potential
      with the above graceful property.    
      We wish to make it clear that our purpose is this, not
      the study of a decoherence process. Decoherence touches
      us only insofar as the result it gives have provided
      the inspiration for this note. 
 \par
      For our purposes we require a very simple
      model as described below.
      We believe that the final result will be to add
      something to a genuinely interesting subject, rather than
      just to present a textbook exercise.

 \vskip .9 true cm
 \noindent{\bf 2. A VERY SIMPLE MODEL }
 \vskip .6 true cm
 \par
      The model consists of a closed one-dimensional system of 
      $N$ noninteracting distinguishable particles with position
      coordinates $x_1$, $x_2$, $\cdots, x_N$  in a infinitely deep
      square potential well $V(x)$.
      This is actually a one-dimensional ideal gas in a container.
      The value of $V(x)$ at the bottom of the well
      is taken as the origin of the energy scale.
      This region of zero potential occupies the
      interval $(0, L)$  of the $x$ axis; it is bounded on both
      sides by infinite potential barriers.
 \par
      We will focus our attention on the above case,
      but will also consider infinitely deep
      potential wells $U(x)$ represented by an analytical
      and positive definite function which 
      approach the form of $V(x)$ in some limits.
      One candidate for $U(x)$ could be:
      \begin{equation}\label{e1}
      U(x) = [ \, 2 \, (x - L/2) / L \, ]^{2m}
      \end{equation}
      with $m$ a positive integer as large as we like
      it to be.
\par
      Apart from the major simplification of no interaction between
      particles, a further simplification we adopt is to consider the
      particles as spinless and distinguishable.
      Each single particle must be identified by its own mass
      to achieve this, all the masses being different.
      However, this would introduce a slight but nevertheless
      unnecessary complication, and we therefore consider
      particles as having the same
      mass $\mu$ but being of different colours:
      the visible spectrum does actually offer a large
      number of possibilities.
 \par
      Let us consider the energy eigenvalues problem
      for a single particle in the infinitely deep square
      potential well.
      The simple and very known result is the set of eigenfunctions:
      \begin{equation}\label{e2}
       \varphi_n(x) = \sqrt{2/L}  \, \sin(n \pi x/L) \qquad
        (n = 1, 2, \cdots, \infty)
      \end{equation}
      which vanishes at extremes $0$ and $L$ and are zero
      outside the well.
      These correspond to the discrete energy values:
      \begin{equation}\label{e3}
       E_n =  (n \pi \hbar)^2 / (2 \mu L^2)
      \end{equation}
      and define a vector space for each individual
      component.
      A generic state of the system can be expressed using
      the tensor product space as:
      \begin{equation}\label{e4}
     \Psi(x_1, x_2, \cdots, x_N)= \sum_{n_1, n_2, \cdots, n_N} 
     c_{n_1, n_2,\cdots, n_N} \, \varphi_{n_1}(x_1) \varphi_{n_2}(x_2)
      \cdots \varphi_{n_N}(x_N)  \quad .
      \end{equation}
      Below we consider the factorized state vectors:
      \begin{equation}\label{e5}
       \Phi(x_1, x_2, \cdots, x_N )= \prod_{i=1}^N  \psi_i(x_i)
      \end{equation}
      with reference to the system subject to the potential
      $U(x)$ and require that the average energy of each particle
      is limited by a common bound $W$:
      \begin{equation}\label{e6}
     <\psi_i(x_i)|\, { \widehat{p_i^2} \over {2 \mu}}  + U(x_i) \, 
      |\psi_i(x_i)> = { <\widehat{p_i^2}> \over {2 \mu}}+<U(x_i)> \, \leq W
      \end{equation}
      where $\widehat{p}_i$ is the momentum operator of
      the $i^{th}$ particle.
      Since the average potential energy is positive, it follows
      that $ <\widehat{p_i^2}> \, \leq 2 \mu W $ and clearly
      the mean square deviation of the momentum
      $ <\widehat{p_i^2}>-<\widehat{p_i}>^2  \leq 2 \mu W $.
      Now indicating with $\sigma_i = < x_i^2>-<x_i>^2$
      the mean square deviation for the position of the $i^{th}$
      particle the uncertainty relation imply that
      $\sigma_i \geq \hbar^2 / (8 \mu W)$ for each $i$. This will 
      be valid still in the limiting case of infinitely
      deep square well.      
      
 \vskip .9 true cm
 \noindent{\bf 3. THE REDUCED MATRIX FOR THE CENTRE OF MASS }
 \vskip .6 true cm
 \par
      As we are interested in the collective behaviour
      of the system, we will use the position co-ordinate
      of the centre of mass:
  	\begin{displaymath}
    X = ( x_1 + x_2 + \cdots + x_N ) / N
  	\end{displaymath}
      and the coordinates  $\xi_i$ of the particles with respect
      to their centre of mass:   $ x_i = X + \xi_i$.
      The $\xi_i$  are not independent, since they must satisfy
      the relation $\sum_{i=1}^N \xi_i = 0$.
      We will consider as independent variables $X$ together with 
      the first $N-1$  relative co-ordinates,
      i$.$e.\ $\xi_1, \xi_2, \cdots, \xi_{N-1}$.
      The last relative co-ordinate
      is expressed as $\xi_N = -\sum_{i=1}^{N-1} \xi_i$.
      The Jacobian determinant of the
      transformation is equal to $N$.
      We will write the integration volume element for the new
      variables as $dV = dV_c \, dV_b$ , where $dV_c = dX$ refers to
      the centre of mass and 
      $dV_b = N d\xi_1 \, d\xi_2 \, \cdots \,  d\xi_{N-1}$
      to the internal degrees of freedom.
      The trace operations in which the continuous spatial reduced
      matrix of the centre of mass may be involved will therefore
      be executed using the integration element $dX$.
 \par
      We now consider a state vector of the system,
      $\Psi(x_1, x_2, \cdots, x_N)$. Its projection $<X'|\Psi>$ on
      a position eigenstate $|X'>$ of the centre of mass is simply 
      $\Psi(X+\xi_1, X+\xi_2, \cdots, X+\xi_{N-1}, X-\sum_{i=1}^{N-1} \xi_i)$.
      The reduced density matrix elements we are interested in
       are obtained by taking the partial
      trace of $<X'| \Psi><~\Psi| X>$ over the microscopic
      degrees of freedom:
      \begin{equation}\label{e7}
      \varrho_{X'X} = \int \Psi^\ast (X+\xi_1, \cdots, X-\sum_{i=1}^{N-1}
       \xi_i) \Psi (X'+\xi_1, \cdots, X'-\sum_{i=1}^{N-1} \xi_i) \,  dV_b 
      \end{equation}
      or taking the relation $\sum_{i=1}^N \xi_i=0$      
      into account with the aid of a delta function:
      \begin{equation}\label{e8}
      \varrho_{X'X} = N \int \delta(\sum_{i=1}^N \xi_i) \Psi^\ast 
       (X+\xi_1, \cdots, X+\xi_N) \Psi (X'+\xi_1, \cdots, X'+\xi_N) \,
       d\xi_1 \, \cdots  d\xi_N 
      \end{equation}

 \vskip .9 true cm
 \noindent{ \bf 4. THE OFF DIAGONAL ELEMENTS }
 \vskip .6 true cm
 \par
      We will now look at the off diagonal elements $\varrho_{X'X}$.
      In the first place, we note that these are limited to the
      range $(0, L)$ of $X$ and $X'$.
      This is clear from observing Eq.~(\ref{e8}) and remembering 
      that the wavefunctions of the system are zero when the
      position coordinates  $x_1$, $x_2$, $\cdots, x_N$ are outside
      this range.
      We may also note that, for the same reason, only the
      integration region for which both inequalities
      $-X < \xi_i < L - X$ and $-X' < \xi_i < L - X'$ are satisfied
      for each $i$ may contribute to the integral in Eq.~(\ref{e8})
      which defines the matrix elements. 
      Let $X' > X$ and $\Delta = X'- X$. 
      Therefore, indicating as $\Omega$
      the volume spanned by the values of the $\xi_i$ in the intervals
      $-X \leq \xi_i \leq L- X'$ and using the expression in
      Eq.~(\ref{e4}) for the generic state vector:
      \begin{equation}\label{e9}
     \varrho_{X'X}= N \sum_{n_1, \cdots, n_N}\sum_{m_1, \cdots, 
     m_N} c^\ast_{n_1, \cdots, n_N} \, c_{m_1, \cdots, m_N} \,
      I_{n_1, \cdots, n_N; m_1 \cdots, m_N}
      \end{equation}
      where
      \begin{equation}\label{e10}
       I_{n_1, \cdots, n_N; m_1, \cdots, m_N}=\int \delta(\sum_{i=1}^N
    \xi_i) \prod_{i=1}^N \varphi^\ast_{n_i}(X + \xi_i) \prod_{j=1}^N
  \varphi_{m_j}(X' + \xi_j) \, d\Omega \quad .
      \end{equation}
      Applying the Schwartz inequality:
      \begin{equation}\label{e11}
      | I_{n_1, \cdots, n_N; m_1, \cdots, m_N}|^2 \leq 
    \int \delta(\sum_{i=1}^N
    \xi_i) \prod_{i=1}^N |\varphi_{n_i}(X+ \xi_i)|^2 \, d\Omega
      \int \prod_{j=1}^N |\varphi_{m_j}(X'+ \xi_j)|^2 \, d\Omega
      \end{equation}
      Since the term to be integrated in the first factor
      is a definite positive
      quantity, we can remove the delta function which
      restricts the integration region and obtain a fortiori:
      \begin{equation}\label{e12}
      | I_{n_1, \cdots, n_N; m_1, \cdots, m_N}|^2 \leq 
    \prod_{i=1}^N \int^{L-X'}_{-X} 
     |\varphi_{n_i}(X+\xi_i)|^2 \, d\xi_i
    \prod_{j=1}^N \int^{L-X'}_{-X}  |\varphi_{m_j}(X'+\xi_j)|^2 \, d\xi_j
      \end{equation}
      If the integration variables are changed
      the previous expression may be rewritten as:
      \begin{equation}\label{e13}
      | I_{n_1, \cdots, n_N; m_1, \cdots, m_N}|^2 \leq 
    \prod_{i=1}^N \int^{L-\Delta}_0 
     |\varphi_{n_i}(x_i)|^2 \, dx_i
    \prod_{j=1}^N \int^L_{\Delta}  |\varphi_{m_j}(x_j)|^2 \, dx_j
    \quad .
      \end{equation}
      Now if we use $ \delta_{\Delta}$  to indicate the maximum value
      with respect to the index $n$ of the quantity:
  	\begin{displaymath}
     \int^{L-\Delta}_0  |\varphi_n(x)|^2 \, dx =
     \int^L_{\Delta}  |\varphi_n(x)|^2 \, dx =
       1 -\int^{\Delta}_0  |\varphi_n(x)|^2 \, dx 
  	\end{displaymath}
      where we used the simmetry properties of the $\varphi_n(x)$
      and the fact that the same wavefunctions are
     normalized, we finally have:
     \begin{equation}\label{e14}
      | I_{n_1, \cdots, n_N; m_1, \cdots, m_N}| \leq 
      \delta^N_{\Delta}  \quad .
      \end{equation}
      To return to the  Eq.~(\ref{e9}) above and using
      the triangular inequality we finally obtain:
     \begin{equation}\label{e15}
    |\varrho_{XX'}| \leq N \, S \, \delta^N_{\Delta} 
      \end{equation}
      where $S$ indicates the double sum: 
  	\begin{displaymath}
        \sum_{n_1, \cdots, n_N} \sum_{m_1, \cdots, m_N}
       |c_{n_1, \cdots, n_N}| \ |c_{m_1, \cdots, m_N}|=
       ( \sum_{n_1, \cdots, n_N} |c_{n_1, \cdots, n_N}|)^2 \quad .
  	\end{displaymath}

      Quantity  $ \delta_{\Delta}$ is clearly less than one
      for each finite value of $\Delta$.
      Therefore, if the value of $S$ is given by a polynomial
      in $N$, the non-diagonal matrix elements tend to zero in
      the limit of large $N$. 
      The results obtained refer to those matrix
      elements for which $ X' \geq X $, but may also be
      extended immediately to the other half, remembering
      that our matrix is hermitean.
 \par
      If only a number $l$ of terms appear
      in the expression of the generic state vector, one can easily
      see that, given the normalisation constraint, the maximum
      possible value for $S$ is $l$.
      Therefore, the condition required to have a corresponding
      spatial reduced density matrix for the centre of mass which
      is diagonal is clearly satisfied by the fundamental state,
      by the low-lying energy eigenstates and by their 
      superpositions.
 \par
      In the case of a factorized state vector as in Eq.~(\ref{e5}),
      by applying the Schwartz inequality once again and after some
      calculations like above, we obtain:
      \begin{equation}\label{e16}
      |\varrho_{X'X}|^2 \leq N^2 \prod_{i=1}^N \left ( 1 - \int_{L-\Delta}^L
      |\psi_i(x_i)|^2 \, dx_i \right ) \, \, \prod_{j=1}^N \left ( 
      1 - \int_{0}^\Delta  |\psi_j(x_j)|^2 \, dx_j \right )  \quad .
      \end{equation}
      If we exclude very unusual forms of the $\psi_i(\xi_i)$
      functions, all factors on the right-hand side
      are less than unity and their product therefore
      decreases rapidly as $N$ increases.
      To be more precise, we assume that the particles have
      the same individual wave function. In which case:
     \begin{equation}\label{e17}
       |\varrho_{X'X}| \leq N a_{\Delta}^{N/2} b_{\Delta}^{N/2}
      \end{equation}
      with $a_{\Delta}$, $b_{\Delta}$  positive quantities
      less than one for any finite $\Delta$ and the off-diagonal matrix
      elements exponentially tending to zero as $N$ increases.

 \vskip .9 true cm
 \noindent{\bf 5. THE DIAGONAL ELEMENTS FOR FACTORIZED STATES}
 \vskip .6 true cm
 \par

      We will now examine the diagonal elements of the reduced
      matrix, with the reference only to a factorized state vector
      as indicated in Eq.~(\ref{e4}), as this is the only case
      in which we can obtain specific results.
      Using Eq.~(\ref{e8})  and the Fourier representation of
      the delta function, we have:
     \begin {equation}\label{e18}
      \varrho_{XX} = \frac {N}{2 \pi} \int dk \prod_{i=1}^N \int 
      e^{i k \xi_i} \psi_i^\ast (X+\xi_i) \psi_i (X+\xi_i) \, d\xi_i
      \end{equation}
      which may be rewritten as:
      \begin{equation}\label{e19}
      \varrho_{XX} = \frac {N}{2 \pi} \int dk e^{-i k N X}
      \prod_{i=1}^N \int e^{i k x_i}
       \psi_i^\ast (x_i) \psi_i (x_i) \, dx_i \quad .
      \end{equation}
      Apart from the factor $N$ (since we are using the differential
      element $dX$ instead of $dY$ here), this is nothing other than
      the expression for the probability density
      distribution of the sum  $NX = Y = x_1 + x_2 + ... + x_N$.
      We could have written this directly using the
      characteristic functions method by observing that the $x_i$
      are distributed in accordance with the $N$ independent probability
      densities $F_i(x_i) = \psi_i^\ast(x_i) \psi_i(x_i)$.
 \par
      We will now consider the second central moments
      $\sigma_i = < (x_i-<x_i>)^2>$
      and the sum $s_N= \sum_{i=1}^N \sigma_i$.
      If we require that the average energy $<E_i> \leq W$
      for each particle, it follows as noted that the second central
      moments are limited by a common bound independent of $N$.
      This means $\sigma_i \geq \alpha$ were $\alpha$ is a positive
      constant.
  \par
      In the case of the infinitely deep square well,
      since the $F_i(x_i)$ are carried by a finite interval,
      the only requirement that each $\sigma_i \geq \alpha$  
      is sufficient to satisfy the Lindeberg condition\cite{fel}
      for the Central Limit Theorem (CLT).
      Therefore, under the conditions indicated, the distribution
      of the diagonal elements $\varrho_{XX}$ will be well
      expressed for a very large $N$ by the Gaussian:
      \begin{equation}\label{e20}
      \sqrt{\frac {N} {2 \pi <\sigma>}} \, e^{-N (X -<X>)^2/ 2 <\sigma>}
      \end{equation}
      where $<\sigma>$ indicates the quantity  $s_N/N$
      and $<X> = \sum_{i=1}^N <x_i>/N$ is the average
      value of the centre of mass position.
      $<X>$ and $<\sigma>$ will depend generally on time,
      but the Gaussian form for large $N$ will in any case 
      be guaranteed since the conditions 
      for the CLT are satisfied independently of time.
      Moreover, as the wave
      functions are zero outside the interval $(0, L)$, 
      the particles mean square deviations are also limited from
      above. We certainly have $\sigma_i \leq L^2$
      for each particle. Therefore $<\sigma> = \sum_{i=1}^N \sigma_i/N$
      remains limited and $<\sigma>/N$ goes to zero as $N$ increase.
      On the time dependence of the central value $<X>$
      and of the variance it can be said that this
      dependence will be of the periodic type, since the
      Bohr frequencies involved are whole multiples of a basic
      frequency, as can be seen from Eq.~(\ref{e3})
 \par 
      In the case of the potential $U(x)$, we once again find
      that the distribution of diagonal matrix elements follows
      the normal law if in addition to the requirement that
      the average energy of each particle is limited (which
      means that their second central moments are limited as well) we
      also consider the third absolute central
      moments $\nu_i = < |\, x_i-<x_i>|^3 >$
      and make the physically reasonable assumption\cite{teg2}
      that $\nu_i \leq \beta$ for some positive constant $\beta$.
      Indeed in this case the Liapunov conditions for the CLM\cite{ren}
      are satisfied.

 \vskip .9 true cm
 \noindent{\bf 6. FINAL REMARKS}
 \vskip .6 true cm
 \par

      We have considered the reduced density matrix of the 
      centre of mass on position basis corresponding to the
      pure states of a one-dimensional
      system of non-interacting distinguishable particles in an
      infinitely deep square well potential.
      By expressing the generic state vector in terms of energy 
      eigenstates, a sufficient condition was identified for
      the matrix in question to be diagonal. The condition is
      that the quantity $S$ defined in the text 
      is limited by a polynomial in the number $N$ of particles.
      This is clearly satisfied by the fundamental state, by the
      low-lying energy eigenstates and by their superpositions.
      Therefore, when the system is in one of these pure states,
      or others for which the condition on $S$ is verified,
      we cannot experience spatial superpositions of the
      centre of mass, its spatial reduced density matrix can
      be thought of in terms of classical probability
      and to conclude the position will appear as a preferred
      basis for the collective variable. However we are
      cautious about this last conclusion, since non-diagonal
      elements of "small" subsystems are known to be small
      in an arbitrarily chosen basis. There even exist
      theorems for the most probable value of entropy of
      such systems\cite{pag}.
 \par
      Since we expressed the generic state vector by means
      of energy eigenstates the coefficients of the expansion
      depend on time only by way of a phase factor, and
      as $S$ depends on their moduli when the condition
      required for this quantity is satisfied,
      it is satisfied permanently.
      The above states for our system of non-interacting
      particles do not therefore appear to be the
      final result of a decoherence process.
      Nevertheless, if we introduce an interaction between
      the particles ( implying a time dependence for the
      moduli of our expansion coefficients ) we could think of the
      possible states which evolve fulfilling at the end
      the condition on $S$
      as the result of a process of decoherence, considering
      our gas as consisting formally of a "collective system"
      described by the centre of mass and an ( internal )
      environment described by the microscopic degrees of freedom.
      Coupling between the two formal systems is guaranteed by
      the external potential.
 \par
      We also considered the
      case in which the system's state vector is factorized in
      the individual wave functions of its components.
      This state vector is a long way from being generic, but
      this assumption is natural if there is no interparticle
      interaction.
      Here too there are indications that the corresponding
      spatial reduced density matrix of the centre of mass
      is diagonal.
      Moreover, if we require the average energy
      of each particle to be less than a common value,
      the conditions for the CLT are satisfied and the diagonal
      elements of the matrix considered are well expressed by
      a Gaussian distribution
      as $N$ increase. The Gaussian
      parameters are time dependent, but the mean square
      deviation decreases in any case as $1/N$. This is
      consistent with a recent work in which we discussed
      the localisation properties of a macroscopic body with
      low total average energy\cite{car}.
      Factorised states of such a type appear for
      any practical purpose as localised states of the
      centre of mass. Each of them appears potentially to
      be the result of successive spontaneous random localisations
      of the individual particles, as contemplated by theories
      of the GRW type.
      An appropriate classical probability distribution of similar
      states localised at various values of the position can
      at a given moment mimic the distribution of diagonal elements
      of the reduced matrix corresponding to the states 
      considered before. But the meaning of the two cases
      is very different\cite{des}.
     
 \par
      To conclude, we must remember that
      some time ago, the example of a macroscopic ball bouncing
      elastically between two parallel walls positioned normally
      to the $x$ axis was used to demonstrate the difficulties
      encountered in describing a similar situation with quantum
      mechanics\cite{jam}. The gas in our example cannot be considered
      a compact body like a ball, but we agree that it is macroscopic.
      If this is described by a factorised state vector like those
      indicated the centre of mass position is distributed
      according to a well localised Gaussian, whose central value
      moves with periodic motion between the two walls.
      However we do not claim that this distribution, which
      is a general consequence of the statistical independence 
      of the particles, represents classical behaviour which
      would require a Gaussian distribution about a time-dependent
      centre for each wave function in the ensemble.
      This could be expected only for a bound state describing
      a macroscopic object, not for a gas.

       \vskip 1.4 true cm

      \end{document}